\documentclass[12pt]{article}
\newcommand {\citeAY} [1] {\citeNP {#1}}%
\newcommand {\citeAPY}[1] {\citeN  {#1}}%
\usepackage{xchicago}
\renewcommand {\showoriginalref}[1]{}
\renewcommand {\showCODEN}[1]{}
\renewcommand {\showISSN}[1]{}
\renewcommand {\showMR}[3]{}

\newcommand\eq[1] {(\ref{#1})}

\newcommand\fig[1] {\ref{fig:#1}}

\newcommand\labfig[1] {\label{fig:#1}}

\newcommand{\bfm}[1]{\mbox{\boldmath ${#1}$}}
\newcommand{\nonum}{\nonumber \\}
\newcommand{\beqa}{\begin{eqnarray}}
\newcommand{\eeqa}[1]{\label{#1}\end{eqnarray}}
\newcommand{\beq}{\begin{equation}}
\newcommand{\eeq}[1]{\label{#1}\end{equation}}
\newcommand{\Grad}{\nabla}
\newcommand{\Div}{\nabla \cdot}
\newcommand{\Curl}{\nabla \times}

\newcommand{\Md}{\partial}

\newcommand{\Ga}{\alpha}
\newcommand{\Gb}{\beta}
\newcommand{\Gd}{\delta}

\newcommand{\Gve}{\varepsilon}

\newcommand{\Gk}{\kappa}

\newcommand{\Gn}{\eta}
\newcommand{\Gm}{\mu}
\newcommand{\Gv}{\nu}

\newcommand{\Gr}{\rho}

\newcommand{\Gs}{\sigma}

\newcommand{\Go}{\omega}

\newcommand{\GD}{\Delta}

\newcommand{\GO}{\Omega}

\newcommand{\BGv}{\bfm\nu}

\newcommand{\BGr}{\bfm\rho}

\newcommand{\BGS}{\bfm\Sigma}

\def\Ba{{\bf a}}

\def\Bf{{\bf f}}

\def\Bj{{\bf j}}

\def\Bn{{\bf n}}

\def\Bt{{\bf t}}
\def\Bu{{\bf u}}
\def\Bv{{\bf v}}

\def\Bx{{\bf x}}

\def\BA{{\bf A}}
\def\BB{{\bf B}}
\def\BC{{\bf C}}
\def\BD{{\bf D}}
\def\BE{{\bf E}}
\def\BF{{\bf F}}

\def\BH{{\bf H}}
\def\BI{{\bf I}}
\def\BJ{{\bf J}}
\def\BK{{\bf K}}

\def\BM{{\bf M}}

\def\BP{{\bf P}}

\def\BR{{\bf R}}

\def\BV{{\bf V}}
\def\BW{{\bf W}}

\def \ba {\begin{array}}
\def \ea {\end{array}}
\newtheorem {Thm} {Theorem} [section]

\newtheorem {Adef} [Thm] {Definition}

\newtheorem {Arem} [Thm] {Remark}

\newtheorem {Aexa} [Thm] {Example}

\newtheorem {Anot} [Thm] {Notation}

\def \refe #1.{(\ref{#1})}
\def \reff #1.{figure~\ref{#1}}
\def \refs #1.{section~\ref{#1}}
\def \refss #1.{subsection~\ref{#1}}
\def \refD #1.{Definition~\ref{#1}}
\def \refT #1.{Theorem~\ref{#1}}
\def \refL #1.{Lemma~\ref{#1}}
\def \refC #1.{Corollary~\ref{#1}}
\def \refP #1.{Proposition~\ref{#1}}
\def \refR #1.{Remark~\ref{#1}}
\def \refE #1.{Example~\ref{#1}}
\def \refN #1.{Notation~\ref{#1}}

\usepackage{graphics}
\usepackage{amssymb}
\begin{document}
\vspace{-1in}
\title{Electromagnetic Circuits}
\author{Graeme W. Milton\\
\small{Department of Mathematics, University of Utah, Salt Lake City UT 84112, USA}\\
Pierre Seppecher\\ 
\small{Institut de Math\'ematiques de Toulon},\\
\small{Universit\'e de Toulon et du Var, BP 132-83957 La Garde Cedex, France}}
\date{}
\maketitle
\begin{abstract}
The electromagnetic analog of an elastic spring-mass network is constructed.
These electromagnetic circuits offer the promise of manipulating
electromagnetic fields in new ways, and linear electrical circuits 
correspond to a subclass of them. They consist of thin triangular magnetic components
joined at the edges by cylindrical dielectric components. (There are  
also dual electromagnetic circuits 
consisting of thin triangular dielectric components joined at the edges by cylindrical magnetic components.)
Some of the edges 
can be terminal edges to which electric fields are applied. 
The response is measured in terms of the real or virtual free currents 
that are associated with the terminal edges. The relation 
between the terminal electric fields and the terminal free currents is governed
by a symmetric complex matrix $\BW$. In the case where all the terminal edges are disjoint,
and the frequency is fixed, a complete characterization is obtained of all possible response matrices $\BW$
both in the lossless and lossy cases. This is done by introducing a subclass
of electromagnetic circuits, called electromagnetic ladder networks.
It is shown that an electromagnetic ladder network,
structured as a cubic network, can have a macroscopic electromagnetic 
continuum response which is non-Maxwellian, and novel.
\end{abstract}
\vskip2mm

\section{Introduction}
\setcounter{equation}{0}

In this paper we introduce a new type of electrical circuit, called an
electromagnetic circuit, which has the potential at a fixed frequency
for providing new and easily analyzable
ways of manipulating electromagnetic fields beyond those provided
by electrical circuits, photonic circuits,
optical lenses, waveguides, photonic crystals, 
and transformation optics. To construct the electromagnetic circuit
we draw upon analogs
between electromagnetism, elastodynamics and acoustics. 
Analogs between electromagnetism and elastodynamics have a long
history [see, e.g. \citeAPY{Silva:2007:RMA} and references
therein] and can be understood
from the viewpoint of differential geometry (\citeAY{Oziewicz:1994:CFT}).
It is easy to
see the connection from the underlying partial differential equations
when they are written in a form which emphasizes the similarity.
Considering for simplicity a locally isotropic medium, Maxwell's equations
at fixed frequency $\Go$ take the form
\beq \BD=\Gve\BE,\quad \BB=\Gm\BH,\quad \Curl\BE=i\Go\BB, \quad
\Curl\BH=\Bj-i\Go\BD,
\eeq{0.1a}
where $\BD(\Bx)$, $\BE(\Bx)$, $\BB(\Bx)$, $\BH(\Bx)$, and $\Bj(\Bx)$
are the complex electric displacement field, electric field,
magnetic induction, magnetic field, and free current
(the physical fields are the real parts of $e^{-i\Go t}\BD$, 
$e^{-i\Go t}\BE$, $e^{-i\Go t}\BB$, $e^{-i\Go t}\BH$, and $e^{-i\Go t}\Bj$
where $t$ is time)
and $\Gve(\Bx,\Go)$ and $\Gm(\Bx,\Go)$ are the complex electric permittivity 
and complex magnetic permeability.  Here the free current $\Bj(\Bx)$ 
may represent a single frequency component of a time varying ion beam current, 
or a time varying current generated by an electrochemical potential. It
does not include conduction currents $\Gs\BE$,
where $\Gs$ is the conductivity, that instead are included in the term 
$\BD=\Gve\BE$, through the imaginary part of $\Gve(\Bx,\Go)$. (It is difficult, if not impossible, to distinguish oscillating displacement currents from 
oscillating conduction currents.) The elimination
of $\BD$, $\BB$ and $\BH$ leads to the form (\citeAY{Milton:2006:CEM}), 
\beq \frac{\Md}{\Md x_p}\left( C_{pqrs}\frac{\Md E_s}{\Md x_r}\right)+i\Go j_q
=-\Go^2\Gve E_q,
\eeq{0.1}
where
\beq C_{pqrs}=e_{pqm}e_{rs m}/\Gm,
\eeq{0.2}
and $e_{pqm}=1$ $(-1)$ if $(p,q,m)$
is an even (odd) permutation of (1,2,3) and is zero otherwise]. This
is clearly similar to the form of the equations of continuum elastodynamics
\beq \frac{\Md}{\Md x_p}\left( C_{pqrs}\frac{\Md u_s}{\Md x_r}\right)+f_q
=-\Go^2\Gr u_q,
\eeq{0.3}
in which $\Bu(\Bx)$ and $\Bf(\Bx)$ are the complex displacement field,
and body force (the physical fields are $\widetilde{\Bu}=(e^{-i\Go t}\Bu)'$
and $\widetilde{\Bf}=(e^{-i\Go t}\Bf)'$ where the prime denotes the real 
part) and $\BC(\Bx,\Go)$ is the complex elasticity tensor 
(incorporating viscosity terms through its imaginary part) 
and $\Gr(\Bx,\Go)$ is the density [which, when it is  
the effective density tensor of an isotropic composite material
can be complex and has the same properties as a function of
$\Go$ as $\Gve(\Bx,\Go)$: see \citeAPY{Milton:2007:MNS} 
and references therein.] At low frequencies,
one often has the approximation that 
$\BC(\Bx,\Go) \approx \BC'(\Bx)-i\Go\BGv(\Bx)$ where 
$\BC'(\Bx)$ is the real component of the elasticity tensor, and $\BGv(\Bx)$
is the viscosity tensor, incorporating both bulk and shear viscosities.
Then, upon introducing the velocity
$\widetilde v={\partial \widetilde u}/{\partial t}$, \eq{0.3} reduces to
\beq
\frac{\partial}{\partial x_q}( C'_{pqrs}
\frac{\partial \widetilde{u}_s}{\partial x_r} 
+\Gv_{pqrs} \frac{\partial \widetilde{v}_s}{\partial x_r}) + \widetilde{f}_p
=\rho  \frac{\partial \tilde v_p}{\partial t} 
\eeq{0.3a}
which may be more familiar to readers acquainted with the Kelvin-Voigt
model of viscoelasticity.

As these analogies have been known for a long time it is 
rather amazing that no electromagnetic 
analog of a spring network with masses at the nodes
has been constructed. Our electromagnetic circuits are this
analog. In elastic networks,
as modeled by the continuum construction of figure \fig{0}, 
the density is concentrated at the nodes, 
while the elasticity is concentrated 
along the edges. Everything is surrounded by void with $\BC=0$
and $\BGr=0$. 

\begin{figure}
\vspace{2in}
\hspace{1.0in}
{\resizebox{2.0in}{1.0in}
{\includegraphics[0in,0in][6in,3in]{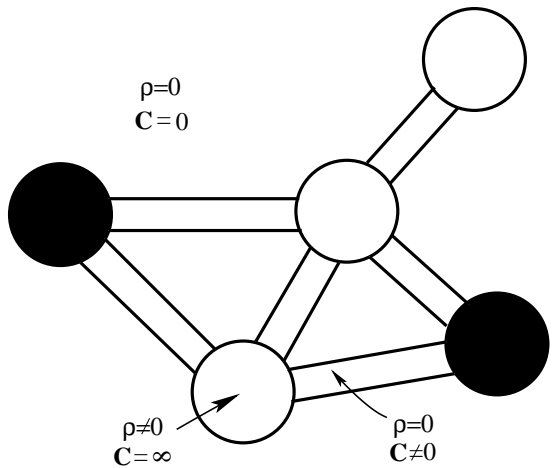}}}
\vspace{0.1in}
\caption{Sketch of a two-terminal discrete elastic network, where the terminal nodes 
are represented by the black circles, and the three internal nodes are represented by the
white circles. In the idealized model these nodes have some mass and are infinitely
stiff. They are are attached by connecting rods (which act as springs) having
no mass and non-zero stiffness. Ideally the nodes should have infinitely small
diameter and the connecting rods should be infinitely
thin, although in practice one then has to worry about buckling which is a non-linear
effect.}
\labfig{0}
\end{figure}

Besides the interest of electromagnetic circuits for providing a new way
of manipulating electromagnetic fields there is also a fundamental reason
for studying them.  It is becoming increasingly clear that the usual 
continuum equations of physics do not apply to composite materials
built from high contrast constituents and having exotic microstructures.
Thus, one can obtain  materials with macroscopic non-Ohmic, possibly non-local,
conducting behavior, even though they conform to Ohm's law at the microscale (\citeAY{Khruslov:1978:ABS}; \citeAY{Briane:1998:HSW}; \citeAY{Briane:1998:HTR}; \citeAY{Briane:2002:HNU},
\citeAY{Camar:2002:CSD}; \citeAY{Cherednichenko:2006:NLH}), materials with
a macroscopic higher order gradient or non-local elastic response even though they are governed by usual linear
elasticity equations at the microscale 
(\citeAY{Pideri:1997:SGM}; \citeAY{Bouchitte:2002:HSE}; \citeAY{Alibert:2003:TMB}; \citeAY{Camar:2003:DCS}),
materials with macroscopic behavior outside that of
continuum elastodynamics even though they are governed by continuum elastodynamics at the microscale (\citeAY{Milton:2007:NMM}), 
and materials with non-Maxwellian macroscopic electromagnetic behavior (\citeAY{Shin:2007:TDE}), even though they
conform to Maxwell's equations at the microscale
[see also \citeAPY{Dubovik:2000:MEE} where other non-Maxwellian 
macroscopic equations are proposed].

One would really like to be able to characterize the possible macroscopic continuum equations that
govern the behavior of materials, including materials with exotic microstructures. A program
for doing this was developed by Camar-Eddine and Seppecher (\citeyearNP{Camar:2002:CSD}, \citeyearNP{Camar:2003:DCS}), and successfully applied to the 
conductivity and elastic equations in the three-dimensional 
static case, assuming the macroscopic behavior
was governed by a single potential or displacement field. The program, in
essence, consists of four steps: first to show that discrete networks 
can be modeled by a continuum construction; second to characterize all 
possible responses of discrete networks allowing for part of the network 
to be hidden; third to find the possible continuum limits of these 
discrete networks; and fourth to show that these possible continuum
behaviors is all that there can be, even when one allows for other, non-network
based, microstructures. For the dynamic case, at fixed frequency, a complete
characterization has been obtained of the possible response matrices of 
multiterminal electrical, acoustic, and elastodynamic networks, both
in two and three dimensions, thus meeting the second goal of the
program in these cases (\citeAY{Milton:2008:RRM}). 
For Maxwell's equations at fixed frequency
our electromagnetic circuits accomplish, in a formal way,
the first goal of the program, and we also make progress towards the
second goal. At this time it is unclear if our electromagnetic
circuits are sufficiently rich in construction that their continuum
limits can model the macroscopic behavior of all other, non-network
based, microstructures, and in particular the question remains open
as to whether the Maxwell equations themselves can be recovered
as a continuum behavior of our electromagnetic circuits. It 
seems clear, however, that many non-Maxwellian continuum behaviors
can be achieved (see the concluding paragraph of the paper).

We emphasize that, besides similarities, 
there are also important differences between Maxwell's equations and the elasticity
equations. For Maxwell's equations the null space of $\BC$ contains all symmetric
matrices, while for elasticity the null space of $\BC$ contains antisymmetric
matrices, which is a space of lower dimension. This manifests itself in the
different boundary conditions: at an interface $\Bu$ is required to be 
continuous, while only the tangential component of $\BE$ is required to
be continuous. In this respect Maxwell's equations have some similarity with
the acoustic equations, which (see \eq{1.1}) take also a form analogous to 
\eq{0.1} or \eq{0.3}, with 
\beq  C_{pqrs}=\Gk\Gd_{pq}\Gd_{rs}, \eeq{0.4} 
and the null space of $\BC$ contains all matrices
which have zero trace, and only the normal component of 
$\Bu$ is required to be continuous at an interface.

There are also linguistic differences when one discusses elasticity
compared to electromagnetism. 
When one wants to study equation (1.2) in a bounded domain, boundary conditions are
needed. A natural condition is to fix the value of $\Bn\cdot\BC\partial \BE/\partial\Bx$ where $\Bn$ is the 
external unit normal to the boundary of the domain. The
analog boundary condition in the elastodynamic case is well known and called the surface
force $\BF$ applied to the medium. Thus we will call the {\it applied surface free current} the
value of $(i\Go)^{-1}\Bn\cdot\BC\partial \BE/\partial\Bx$ on the boundary  and denote it
$\BJ$. (The additional factor of $(i\Go)^{-1}$ is introduced because $i\Go\Bj$ in 
\eq{0.1} plays the role of $\Bf$ in \eq{0.3}.)
This is not a usual way of speaking in the electromagnetic framework as the value
of $(i\Go)^{-1}\Bn\cdot\BC\partial \BE/\partial\Bx$ is nothing else than the tangential
part of the magnetic field $\BH$ at the boundary.
The interest of such a vocabulary appears later. 

When one wants to study equation (1.2) in a domain $\Omega$ which is divided in two
subdomains $\Omega_1$, $\Omega_2$, one has to write jump conditions on the dividing
surface. This condition is the continuity of the tangential part of $\BH$ (the analog of
which is the continuity of the normal part of the stress in elastodynamic framework). 
Alternatively one says in the elastodynamics framework that $\Omega_2$ exerts on $\Omega_1$
a surface force $\BF$ while $\Omega_1$ exerts on $\Omega_2$ the opposite surface force
$-\BF$. This action-reaction law makes the link with the separate study of both subdomains
as it fixes the needed boundary conditions for these studies. 

In a similar way we can say
that $\Omega_2$ exerts on $\Omega_1$ a surface free current $\BJ$ while $\Omega_1$ exerts
on $\Omega_2$ the opposite surface free current $-\BJ$. This way of thinking needs some
practice to become natural and the reader should be aware that this formulation does not
mean, in any way, that there exist actual free currents in the material (just like
action-reaction law does not imply the existence of actual surface forces inside the
domain). However the surface force $\BF$ that $\Omega_2$ exerts on $\Omega_1$  has an
equivalent effect on $\Omega_1$ as a body force $\Bf$ concentrated at the boundary replacing 
the stress field in $\Omega_2$, and similarly the applied surface free current 
$\BJ$ that $\Omega_2$ exerts on $\Omega_1$ has an 
equivalent effect on $\Omega_1$ as a free current $\Bj$  concentrated at the boundary replacing the
$\BH$ field in $\Omega_2$. 

It is well known that when $\Bj=0$ the Maxwell system of equations 
remains unchanged when one interchanges the roles of $\BE$ and
$\BH$ and of $\Gve$ and $\mu$. Therefore
for each electromagnetic circuit discussed here, there is a dual 
magnetoelectric circuit (ME-circuit) obtained
by making these replacements. Instead of speaking about applied free surface electric currents, we 
could speak about applied free surface magnetic monopole currents. These are then truly 
unphysical, but their introduction is again just a device for keeping track
of boundary conditions. 

Throughout the paper we use infinite or zero values of various moduli. From a physical
viewpoint one should think of such moduli as just being positive and
real and extremely large or extremely small.
From a mathematical viewpoint one should think of taking the limit as these
moduli approach infinity or approach zero. Generally values of the permittivity and 
permeability near zero or infinity are difficult to achieve. However, using resonance effects 
(\citeAY{Schelkunoff:1952:ATP}; \citeAY{Pendry:1999:MCE})
very small or very large values which are almost real and positive
may be achieved over a narrow frequency range. The importance of this was recognized
by \citeAPY{Engheta:2005:CEO} and \citeAPY{Engheta:2007:CLN} who realized one could
build nanoscale equivalents of electrical circuits using such materials: a material
with $\Gve$ near zero electrically insulates the circuit, while a material
with $\Gve$ near infinity provides the necessary electrical connections.

From a physical viewpoint perhaps the greatest barrier to the construction 
of electromagnetic circuits is the use of a matrix, which is the
electrodynamic equivalent of a void in elasticity, and
has an extremely large value of the magnetic permeability $\Gm$, and
(although it is not clear it is necessary) 
an extremely small value of the electric
permittivity $\Gve$. (In the case of the dual circuits, one would
need the reverse). In fact it is not necessary that the circuit be 
embedded in a body with these properties, only that a material with
these properties clads the circuit. Also note that the Maxwell 
equations \eq{0.1} remain valid if one
divides $\Gm(\Bx)$ everywhere by a constant $k$ and correspondingly
multiplies $\Bj(\Bx)$ and $\Gve(\Bx)$ by $k$. Therefore it 
should be possible to renormalize
the moduli in the EM-circuit in such a way, that the moduli in the matrix
take more realistic values, perhaps even that of empty space with
$\Gve=\Gm=1$. This is similar to the way a spring-mass network can still
function when embedded in an elastic material provided the springs are
appropriately stiff, the forces sufficiently strong, 
and the masses are sufficiently heavy. 

The objective of this paper is to introduce the concept of EM-circuits
and their basic properties. The approach is formal, but will
hopefully motivate future analytical and numerical work to place
the treatment given here on a firm foundation.

\section{Transverse electric EM-circuits}
\setcounter{equation}{0}
We are interested in the Helmholtz equation
\beq  \Div(1/\Gm\Grad E)=-\Gve\Go^2E, \eeq{1.3}
describing three-dimensional TE electromagnetic wave propagation, where 
$\Go$ is the (fixed) frequency of oscillation,
$\BE(\Bx)=(0,0,E)$ 
is the electric field, $\Gve(\Bx)$ is the electrical permittivity, 
$\Gm(\Bx)$ is the magnetic permeability, and all of these quantities do not
depend on $x_3$. Given the electric field component $E$ the
associated in-plane magnetic field is
\beq \BH=-[i/(\Gm\Go)]\BR_\perp\Grad E, \eeq{1.4}
where
\beq \BR_{\perp}=\pmatrix{0 & -1 \cr
                          1 & 0 \cr}
\eeq{1.5}
is the matrix for an anticlockwise rotation by $90^\circ$ in the plane. When 
in some subregion $\GO$, the moduli are real, positive, and do not 
depend on frequency, the electromagnetic energy stored in $\GO$ is
\beq W(\GO)=\int_{\GO}[\Gm|\BH|^2+\Gve |E|^2]/4.
\eeq{1.5a}
We are only interested in solutions such that $W(\GO)$ remains bounded
in all subregions $\GO$ where the moduli tend to zero or infinity,
remaining real and positive in this process.

The Helmholtz equation \eq{1.3} is mathematically
analogous to the acoustic equation which in two-dimensions reads as
\beq \Div(1/\Gr\Grad P)=-(1/\Gk)\Go^2P, \eeq{1.1}
where $P(\Bx)$ is the pressure, $\Gk(\Bx)$ is the bulk modulus, and $\Gr(\Bx)$ 
is the density. Given the
pressure field $P(\Bx)$ the associated velocity field of the fluid is
\beq \Bv=-[i/(\Gr\Go)]\Grad P, \eeq{1.2}
and when in some subregion $\GO$ the moduli are real, positive,
and do not depend on frequency the time averaged 
elastokinetic energy in $\GO$ is
\beq W(\GO)=\int_{\GO}[\Gr|\Bv|^2+|P|^2/\Gk]/4.
\eeq{1.2a}
(The extra factor of $2$ arises because the physical velocity
and pressure is the real part of $e^{-i\Go t}\Bv$ and $e^{-i\Go t}P$).

To conceive TE electromagnetic circuits we just have to understand how 
discrete acoustic networks are made and
transcribe their structure in terms of electrodynamic quantities.
As illustrated in figure \fig{1} 
we consider a network of channels  
connected by junctions. Each channel 
has parallel sides and contains a segment of 
incompressible, non-viscous, fluid with some constant density $\Gr$ possibly varying from channel to channel, moving in a time harmonic oscillatory 
manner in response to time harmonic pressures at the junctions. 
We define the entire cavity associated with a junction to be the 
cavity at the junction, plus the remaining 
region in the channels not occupied by the incompressible fluid. 
If the junction 
is a terminal so that there
is an open channel leading to it, we insert a segment of massless incompressible fluid
in that channel to keep track of the response of the acoustic network. 

\begin{figure}
\vspace{2in}
\hspace{1.0in}
{\resizebox{2.0in}{1.0in}
{\includegraphics[0in,0in][6in,3in]{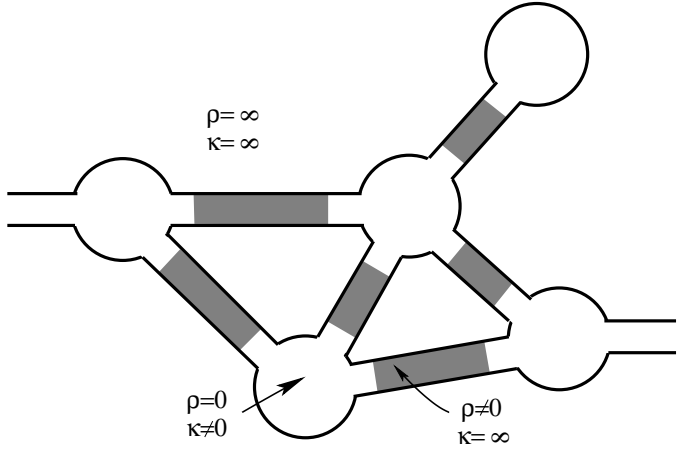}}}
\vspace{0.1in}
\caption{A two-terminal discrete acoustic network. In the idealized model the four cavities 
contain compressible massless fluid, while the grey shaded fluid plugs in the five tubes 
contain incompessible fluid with some mass.}
\labfig{1}
\end{figure}

Each entire cavity contains a compressible, non-viscous, massless fluid with compressibility
possibly varying from cavity to cavity. In this model the compressibility is localized
in the cavities and the mass is localized in the channels between cavities.
The surfaces between the compressible and 
incompressible fluids are assumed to be flat and perpendicular to the channel.
When one conceives an acoustic network, the area outside
the network is assumed to be occupied by a rigid boby (or cladded by rigid tubes). However we can, in a
equivalent way, assume that this area  $\GO_0$ is filled by 
an incompressible fluid having infinite density, i.e. with $\Gk=\Gr=\infty$.
Indeed, the infinite density, and the boundedness of $W(\GO_0)$ 
ensures that the velocity $\Bv$ 
(as one might physically expect) will be zero outside the 
network and consequently that the walls of the network remain fixed. 
The incompressibility ensures that both sides of \eq{1.1} are zero in 
the matrix, without requiring that $P=0$ in the matrix.
Hence the acoustic equation \eq{1.1} desribes the system in the whole space.

In each entire cavity, $\GO_j$, where $1/\Gr$ is infinite (or more precisely
is real positive and approaches infinity), $\Grad P$ 
must be zero, since otherwise
 $\Gr|\Bv|^2=|\Grad P|^2/(\Gr\Go^2)$ would be infinite
and $W(\GO_j)$ would be unbounded. Thus,
as expected, the pressure is constant in each junction region. 
Within each segment 
of incompressible fluid of constant density, both sides of \eq{1.1} must vanish, 
which implies $\Grad^2P=0$ in each
such segment. From the boundary conditions (that $P$ is constant at the
ends of the fluid segment, and at the sides $\Bn\cdot\Grad P=0$, since
$\Bn\cdot\Bv=0$) it follows that the pressure $P$ will be constant
in each cross section normal to the channel, and will vary linearly along the fluid segment.
The fluid velocity $\Bv$ will therefore be constant in the segment, and directed parallel to 
the fluid channel. From \eq{1.2} we see that in a channel joining cavity $i$ and cavity $j$
the fluid velocity in the direction of the channel, from $j$ to $k$ will be 
\beq v_{jk}=-[i/(\Gr_{jk}\Go\ell_{jk})](P_k-P_j), \eeq{1.6}
where $\ell_{jk}$ is the length of the fluid segment, $\Gr_{jk}$ is its density
and $P_j$ and $P_k$ are the complex pressures at junctions $j$ and $k$ respectively. 
This is basically Newton's second law, relating the acceleration of the fluid segment, 
$-i\Go\Bv_{i,j}$, to its mass and the force acting on it. 

In entire cavity $j$, \eq{1.1} and \eq{1.2} imply
\beq \Div \Bv=(i/\Gk)\Go P_j, \eeq{1.7}
which when integrated over the entire cavity implies, by the divergence theorem,
\beq \sum_{k=1}^m hv_{jk}=a_j(i/\Gk_j)\Go P_j,
\eeq{1.8}
where $a_j$ is the area of the entire cavity, $\Gk_i$ is
the bulk modulus of the fluid within it, and
we have assumed that $m$ channels enter the cavity, each with
width $h$ and carrying a fluid segment with velocity
$v_{jk}$. This is essentially Hooke's law, applied to the
compressible fluid occupying the entire cavity. If the cavity 
is a terminal and there is an open channel carrying a current $I_j$
into it, then the relation \eq{1.8} takes the modified form
\beq I_j+\sum_{k=1}^h hv_{jk}=a_j(i/\Gk_j)\Go P_j.
\eeq{1.9}
Given the pressures $P_j$ at the terminal cavities, the equations \eq{1.6} and \eq{1.8} 
provide a discrete set of equations, which can be solved for the pressures 
in the other cavities, the velocities of the fluid plugs in the channels, 
and the currents $I_j$ flowing into the terminal cavities. Assuming that the cavities
are numbered in such a way that the first $n$ are terminals, and the remaining ones
are not, the response of the network is expressed in terms of the linear relation
\beq \BI=\BM\BP \eeq{1.10aa} 
between the set of pressures $\BP=\{P_1,P_2,\ldots,P_n\}$ at the terminals and
the currents $\BI=\{I_1,I_2,\ldots,I_n\}$ flowing into them.

\vskip1cm

Everything carries through to the electromagnetic case where the fields are 
transverse electric (TE).
By comparing \eq{1.1} and \eq{1.2} with \eq{1.3} and \eq{1.4} 
we see that $\Gm$ and $\Gve$ play the role of $\Gr$ and $1/\Gk$; $E$ plays the role
of $P$ and $\BH$ plays the role of $\BR_{\perp}\Bv$ (and is therefore
perpendicular to the channel walls). The channels themselves are now thin 
plates containing a material with $\Gve=0$ and $\Gm\ne 0$. The cavities
are now aligned dielectric cylinders with $\Gve\ne 0$ and $\Gm=0$. The electric
field is constant in each cylinder, which also can be seen directly from the result
of \citeAPY{Silveirinha:2006:TEE} who show
that, for the dual transverse magnetic (TM) problem, cylinders having
$\Gm\ne 0$ and $\Gve=0$ have a constant magnetic field in them. 
It follows that $\BH=0$ in the matrix 
by direct analogy with the acoustic case where $\Bv=0$ in the matrix.
(Although  magnetic fields tend to be concentrated inside a material with 
positive and very large permeability, this concentration refers to the
$\BB$ field and not to the $\BH$ field).

We call such a circuit a transverse electric 
EM-circuit (see figure \fig{2}). Each equation we discussed has its 
analog. For example
\eq{1.9} becomes
\beq I_j+\sum_{k=1}^m hH_{jk}=a_j i\Go\Gve_j E_j, \eeq{1.10}
where  $H_{jk}$ is the value of $\BH$
in the direction perpendicular to the walls of the plate $jk$,
$I_j$ is the line integral of $\BH$ across the open channel,
$\Gve_j$ is the dielectric constant of cylinder $j$ while 
$E_j$ is the electric field in the cylinder $j$.
So the left hand side of \eq{1.10} is the line integral of $\BH$ around the
terminal dielectric cylinder, while the right hand side of \eq{1.10} 
is the total displacement current flowing through the cylinder. 
Thus \eq{1.10} is nothing but Ampere's circuital law (with 
Maxwell's correction). Notice that instead of having an open channel one 
could have a free current $-I_j$ flowing next to the terminal dielectric
cylinder.

The analog of \eq{1.6} is 
\beq  H_{jk}=-[i/(\Gm_{jk}\Go\ell_{jk})](E_k-E_j),
\eeq{1.10a}
where $\Gm_{jk}$ is the permeability of the plate $jk$. This is Faraday's 
law of induction, relating the time derivative of flux of $\BB$
through any rectangle with two opposite sides along the dielectric
cylinders $j$ and $k$, to the line integral of $\BE$ around
this rectangle. Since $\BH$ is constant and perpendicular
to the plate walls, it follows from \eq{1.4} that $\BE$
in the plate depends linearly on $x_1$ and $x_2$ in such a
way that it is constant perpendicular to the plate.

In the particular case
when the cylinder $j$ has zero dielectric constant, i.e. $\Gve_j=0$,
(so that the junction is the analog of a cavity filled with
incompressible fluid) \eq{1.10} becomes
\beq I_j+\sum_{k=1}^m hH_{jk}=0, \eeq{1.11}
where $I_j=0$ if the cylinder is not a terminal cylinder. If all 
cylinders have zero dielectric constant, then we call the circuit 
a transverse electric magnetic circuit (M-circuit). 

\begin{figure}
\vspace{2in}
\hspace{1.0in}
{\resizebox{2.0in}{1.0in}
{\includegraphics[0in,0in][8in,4in]{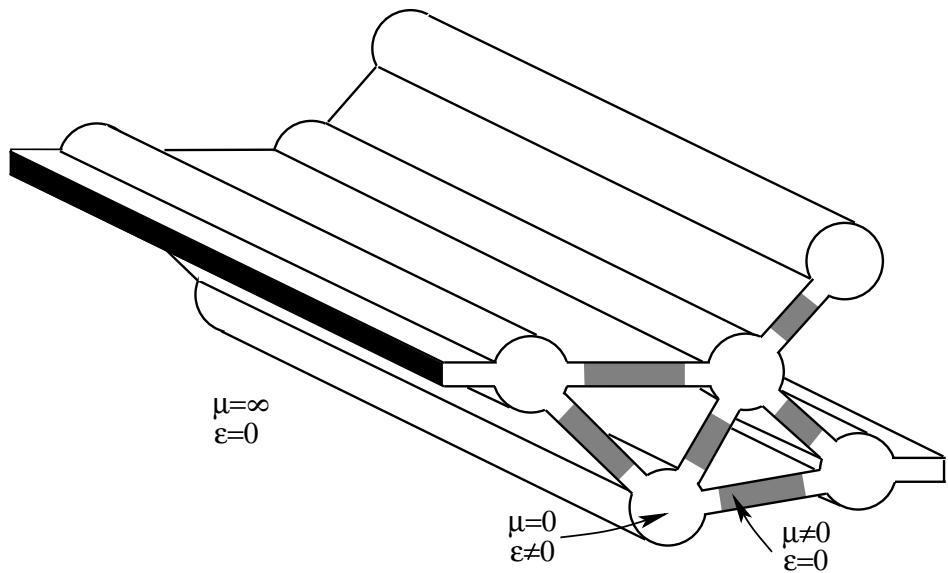}}}
\vspace{0.1in}
\caption{A two-terminal transverse electric EM-circuit 
which is the exact analog
of the discrete acoustic network of Figure 1}
\labfig{2}
\end{figure}
\medskip
It is now important to understand how can these circuits be used and in
particular how they can interact with ordinary materials. The problem is analogous for
connecting an acoustic discrete network to an ordinary acoustic three-dimensional domain. 
Assume that the matrix with $\Gm=\infty$ and $\Gve=0$ only has finite
extent, and is surrounded by space with $\Gm=\Gve=1$, in which
there are TE fields. Also suppose each terminal edge is connected
to the exterior by an open channel, of width h, containing material with
$\Gve=\Gm=0$. 
The external field $E$ (which is the anolog of the pressure) will fix the mean value of
$E_j$ at every open channel mouth $j$. Then the response of the transverse EM circuit 
will determine the values $I_j$, that is of $\BH$ (which is analog to the velocity) at the open channel
mouths. Let $\Omega_0$ denote the region occupied by the circuit plus the remaining matrix.
On the part of the boundary of $\Omega_0$ which corresponds to the matrix we have $\BH=0$. Hence
the $E$-to-tangential value of $\BH$ map (which is equivalent to the Dirichlet to Neumann map) of $\Omega_0$ will
be governed by the response of the circuit. It will be completely different from a pure matrix
(for which the tangential value of $\BH$ vanishes on $\partial \Omega_0$) or from void where $ \Gve=\mu=1$.

Note that the external field $E$ will fix the value of the electrical field at each open
channel mouth in a efficient way if $h$ is large enough. If $h$ is too small this
connection will be weak, and $E$ near each mouth will be strongly affected by 
flux of $\BR_\perp\BH$ (which is analogous to current in the acoustic setting)
through the narrow channel openings. (In a region
near but not too close to each mouth the $E$ field will be like that generated from a 
line source.) However this problem can be corrected by adding at each open
channel mouth a material with $\Gve=\Gm=0$ (see figure \fig{2a}). Hence the value of $E$ at each mouth
will be fixed by the value of $E$ on the 'relatively large' cap boundary, and the flux
of $\BR_\perp\BH$ through the channel will be transferred to the outside of the
cap. This is similar to the way \citeAPY{Silveirinha:2006:TEE} has, for the dual TM problem, 
suggested the use of  materials with $\Gve=0$ for transfering energy through narrow openings.

\begin{figure}
\vspace{2in}
\hspace{1.0in}
{\resizebox{2.0in}{1.0in}
{\includegraphics[0in,0in][4in,2in]{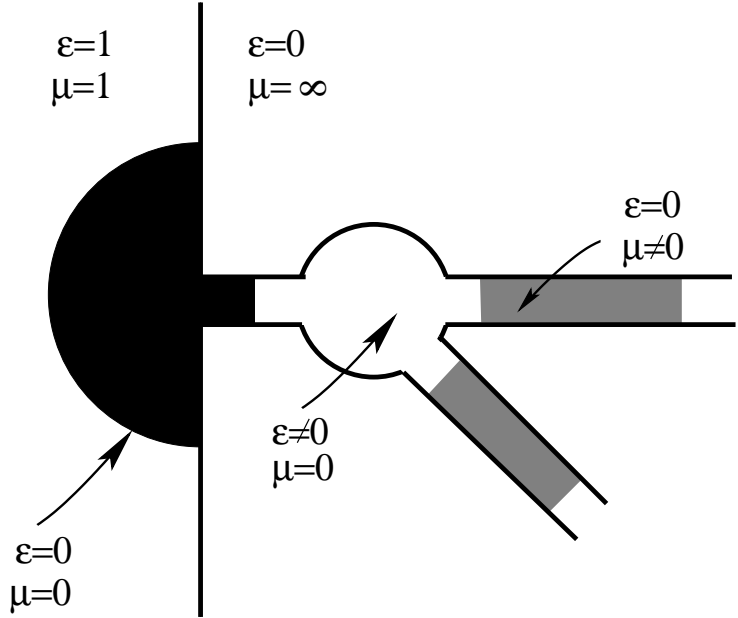}}}
\vspace{0.1in}
\caption{A semicircular plug of material with $\Gve=\Gm=0$ can serve 
to couple the open channels of an transverse electric EM-circuit
with exterior TE fields, allowing the $\BH$ field to be transfered
to the exterior with negligable drop in $E$. }
\labfig{2a}
\end{figure}

\section{Electromagnetic circuits in the general case}
\setcounter{equation}{0}
We need to generalize the EM-circuits to allow for fields
that are not transverse electric. Like in the transverse
electric case the circuit will be composed of plates of 
material having $\Gve=0$ and $\Gm\ne 0$, joined by cylinders of
dielectric material having $\Gve\ne 0$ and $\Gm=0$, embedded in a 
matrix having $\Gm=\infty$  and $\Gve=0$, so that the matrix
is the electrodynamic equivalent of a void in elasticity according
to \eq{0.1}-\eq{0.3}.

The plates and the dielectric cylinders play the physical role in our circuits that
springs and masses play in an elastic network, despite the fact that they are completely
different geometrical objects. The assumption that the electromagnetic
energy density in the matrix remains bounded as $\Gm\to\infty$
and $\Gve\to 0$ when the moduli $\Gm$ and $\Gve$ 
are positive and independent of
frequency, again implies that $\BH=\BD=0$ in the matrix. 
(Note that if $\BH=0$ in the matrix then necessarily
$\BD=i\Curl\BH/\Go$ is also zero).

We emphasize that when $\Gve$ and $\Gm$ are real and positive in the matrix
and $\Gm$ is very large, while $\Gve$ is very small
then there certainly exist (high energy)
solutions where $\BH$ in the matrix is not small: after all an electromagnetic
wave could propagate there, and its amplitude scaled as one desires.
However, we believe (and this needs
to be rigorously verified) that the solutions in the matrix almost
decouple from the solutions in the electromagnetic circuit
when $\Gm$ is very large and $\Gve$ is very small. This should be 
similar to the way
electromagnetic fields almost decouple at a planar interface
between two non-absorbing media, 1 and 2, for which there is a large mismatch
in the electromagnetic impedances $\Gn_1=\Gm_1/\Gve_1$ 
and $\Gn_2=\Gm_2/\Gve_2$: when $\Gn_1/\Gn_2$ is very large
then a plane electromagnetic wave incident from either side of
the interface will have only a tiny portion of its energy
transmitted. 

Alternatively,
and as kindly suggested to us by a referee, one may assume that in the
matrix the product $\Gve\Gm$ is large and negative.
Then electromagnetic fields in the
matrix will be confined within a small skin depth of the surface
which tends to zero as $\Gve\Gm\to -\infty$, again implying 
that $\BH=\BD=0$ in the limit as $\Gm\to\infty$ and $\Gve\to 0$ in such
a way that $\Gve\Gm\to -\infty$.

Let us now analyze in detail the response of each plate. The 
plate could be polygonal in shape, but for simplicity we use
a basic element which is a very thin triangular prism, of uniform height $h$ containing a material having $\Gve=0$ 
and $\Gm\ne 0$, the top and bottom faces of which are
surrounded by the matrix, as illustrated in figure
\fig{3}. We call this element a magnetic element.
\begin{figure}
\vspace{2in}
\hspace{1.0in}
{\resizebox{2.0in}{1.0in}
{\includegraphics[0in,0in][6in,3in]{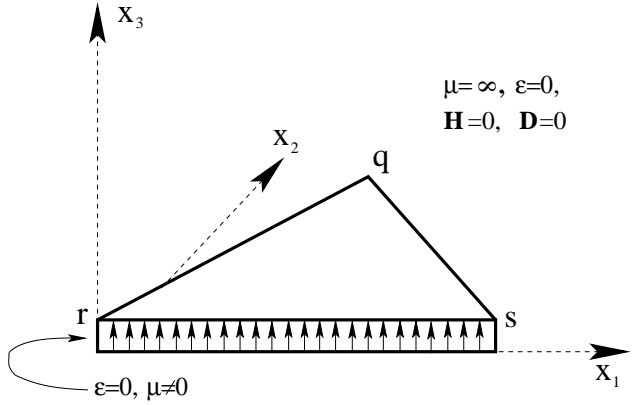}}}
\vspace{0.1in}
\caption{A triangular prism containing material with $\Gve=0$ and $\Gm\ne 0$ is one
basic element of a magnetic or electromagnetic circuit. Here the arrows denote direction
of the magnetic field $\BH$ within the prism.}
\labfig{3}
\end{figure}

Let us choose our coordinate system
so the bottom surface of the prism is at $x_3=0$ and the top surface at $x_3=h$. 
The triangle at the bottom of the prism has vertices
$q$, $r$ and $s$, labeled in an anticlockwise 
order when viewed from above, and edges
$qr$, $rs$ and $sq$. Let $\Gm_{qrs}$ denote the 
constant value of $\Gm$ within
the prism. Since $\Div\BH=0$ and $\Curl\BH=0$ in the prism it follows
that $\BH=\Grad\psi$ where $\GD\psi=0$. Also since the tangential component
of $\BH=\Grad\psi$ is zero at the top and bottom surfaces of the prism, it follows
that $\psi$ is constant on the top and bottom plates: $\psi$ is like the
potential between two closely spaced capacitor plates. Hence $\BH$ is 
essentially constant
within the prism and normal to the top and bottom surfaces,
i.e. $\BH=(0,0,H_3)$ where $H_3$ cannot depend on $x_3$ since
$\Div\BH=0$ within the prism. (In fact $\BH$ will only be approximately constant
due to fringing fields which, however, should become negligible away from the
edges, in the limit as the prism becomes very thin.)

Let $\Bx'=(x_1, x_2)$ denote coordinates in the plane.
Assuming the point $r$ is at $x_1=x_2=0$,
the three edges of the triangle lie along the three lines
\beq \Bx'=\Ga\Bt_{qr}, \quad \Bx'=\Ga\Bt_{rs}, \quad  \Bx'=\Bx'_0+\Ga\Bt_{sq},
\eeq{2.1}
each parameterized by $\Ga$ where $\Bt_{mn}$ is the unit vector directed from
vertex $m$ to vertex $n$, and $\Bx'_0$ is a point along the edge $sq$.
In electromagnetic circuits we
constrain the tangential component of the electric field
to take constant values $E_{qr}\Bt_{qr}$, $E_{rs}\Bt_{rs}$, 
and $E_{sq}\Bt_{sq}$ along the three sides $qr$, $rs$ and $sq$ 
of the triangle. (As we will see later the presence of
dielectric cylinders along these edges will allow this constraint to be
satisfied).  Let $\ell_{qr}$, $\ell_{rs}$ and $\ell_{sq}$ denote the 
lengths of the edges $qr$, $rs$ and $sq$. Then Faraday's law of induction
applied to a circuit around the triangle implies
\beq \ell_{qr}E_{qr}+\ell_{rs}E_{rs}+\ell_{sq}E_{sq}=ca_{qrs},
\eeq{2.7}
where $c=i\Go\Gm_{qrs} H_3$ and $a_{qrs}$ is the area of the triangle.

To find an explicit expression for the electric field
in the prism, although it is not clear we need it, 
let us assume
that $\Gve$ in the prism is arbitrarily small, 
but non-zero (and many factors greater than the $\Gve$ in the
matrix which we treat as being zero),
so that in the prism $\Div\BD=0$ implies $\Div\BE=0$.
Since $\BD=0$ in the matrix and $\Div\BD=0$
it follows that $\BD$ and hence $\BE$ are tangential to
the top and bottom surfaces of the prism. 

Having a material with zero permittivity outside the prism 
allows us to have a non-zero $\BE$ field there.
Then the tangential components of $\BE$ can be 
continuous across the top and bottom surfaces of the prism. 
It is not clear that this zero permittivity in the matrix is necessary.
One could instead have $\Gve\ne 0$ and $\BE=0$ outside the prism, with
a concentrated surface $\BB$ current to compensate for
the jump in the tangential component of $\BE$ across the
surface. Such a concentrated surface $\BB$ current should be allowed since
$\Gm=\infty$ in the matrix. (Similarly in an elastic network, it is not 
necessary that the surrounding material have density $\Gr=0$, although that is the case when
the surrounding material is void. If $\Gr$ is non-zero and $\BC$ is close to zero
then only a small boundary layer near the elastic network will move.)
  
Since $\Curl\BE=i\Go\Gm\BH$, we infer that
\beq \BE=(-cx_2/2,cx_1/2,0)+\Grad\phi, 
\eeq{2.0a}
where $\GD\phi=0$ and without loss of generality
one can assume that $\phi=0$ at the origin. The
potential $\phi$ satisfies the
Neuman boundary conditions that $\Bn\cdot\Grad\phi=0$
on the top and bottom surfaces of the prism, 
and Dirichlet boundary conditions on the
sides of the prism (specifying the tangential
value of $\Grad\phi$ around the sides, and
the value $\phi=0$ at the origin determines
the value of $\phi$ along the sides). Thus 
$\phi$ is uniquely determined and a simple 
calculation using \eq{2.1}, the identity
\beq \ell_{qr}\Bt_{qr}+\ell_{rs}\Bt_{rs}+\ell_{sq}\Bt_{sq}=0, \eeq{2.6}
(as follows from the fact that the edges form a triangle)
and the fact that  $a_{qrs}=\ell_{sq}\Bt_{sq}\cdot\BR_\perp\Bx_0/2$ is the area of the triangle
(as can be easily seen by choosing $\Bx'_0$ to be perpendicular to $\Bt_{sq}$ )
shows that the boundary 
conditions are satisfied with
$\Grad\phi=(a_1,a_2,0)$ where $\Ba=(a_1,a_2)$
is constant and determined by 
\beq \Bt_{qr}\cdot\Ba=E_{qr}, \quad \Bt_{rs}\cdot\Ba=E_{rs}.
\eeq{2.0b}
(The condition \eq{2.7} ensures that $\BE\cdot\Bt_{sq}=E_{sq}$
along the edge $sq$).

It is natural to introduce three new variables
\beq V_{qr}=\ell_{qr}E_{qr}, \quad V_{rs}=\ell_{rs}E_{rs}, 
\quad V_{sq}=\ell_{sq}E_{sq},
\eeq{2.8}
which when $H_3=0$ would represent the potential drops along the three edges.
Then \eq{2.7} implies that $H_3$ depends on $V_{qr}$, $V_{rs}$, and $V_{sq}$
only through the sum $V_{qr}+V_{rs}+V_{sq}$. 

In keeping with the vocabulary introduced in the
introduction, the material surrounding the edges
of the basic element exerts total applied surface free currents  
$J_{qr}^s$, $J_{rs}^q$ and $J_{sq}^r$ along the 
edges $qr$, $rs$ and $sq$,
flowing in directions $\Bt_{qr}$, $\Bt_{rs}$, and $\Bt_{sq}$,
where the superscript is kept to signify that the currents are associated with
the triangle $qrs$. (Here total signifies that these are the applied surface free
currents integrated over the width of each edge, but from now on this will be
assumed so we will drop the word total).
In other words, the boundary conditions on the edges of the basic
element are essentially the same as if we completely surrounded the basic element by 
matrix material with $\Gve=0$ and $\Gm\ne 0$ having
$\BH=0$ and inserted these surface free currents along the edges.

These currents are all equal, and 
from Ampere's circuital law applied to a circuit around each edge 
take the value $hH_3$, by virtue of the fact that $\BH$ is constant within the triangular prism. It
may seem superfluous to keep track of the three currents $J_{qr}^s$, 
$J_{rs}^q$ and $J_{sq}^r$ when they are all equal. 
However, consider the analogous elastodynamic framework: to write the 
balance of forces at each node, one introduces the forces that each spring
exerts on each node even though the forces exerted by a spring on its 
two extremity nodes are equal and opposite. Without 
introducing  $J_{qr}^s$, $J_{rs}^q$ and $J_{sq}^r$
it would be difficult to derive an expression
for the response matrix of an general electromagnetic circuit, as we
do in section 5.

Thus we have the relation
\beq\pmatrix{T_{qr}^s \cr T_{rs}^q \cr T_{sq}^r}\equiv 
\pmatrix{i\Go J_{qr}^s \cr i\Go J_{rs}^q \cr i\Go J_{sq}^r}=k_{qrs}\pmatrix{1 & 1 & 1\cr
                                           1 & 1 & 1 \cr
                                           1 & 1 & 1 \cr}
\pmatrix{V_{qr} \cr V_{rs} \cr V_{sq}},
\eeq{2.9}
where $k_{qrs}=h/(a_{qrs}\Gm_{qrs})$. We use the quantities $T$ rather than the free currents $J$ 
so that the matrix entering the above relation is real
and so that the parallel with elastodynamics is maintained, since
$i\Go\Bj$ in \eq{0.1} plays the role of the body force $\Bf$ in \eq{0.3}. 

Also our introduction of the variables $V$ rather than the variables $E$ ensures that the
matrix is symmetric which is desirable since this property will then 
extend to the matrix describing the response of electromagnetic circuits
with many elements. (Again, this is why it is important to introduce
the three currents $J_{qr}^s$, $J_{rs}^q$ and $J_{sq}^r$ rather than
just a single current.) This relation \eq{2.9},
which is essentially Faraday's law of induction,
is the analog of Hooke's law,
\beq \pmatrix{\BF_{r}^s \cr \BF_{s}^r}
=k_{rs}\pmatrix{\Bt_{rs}\otimes\Bt_{rs} & -\Bt_{rs}\otimes\Bt_{rs}  \cr
                -\Bt_{rs}\otimes\Bt_{rs} &  \Bt_{rs}\otimes\Bt_{rs} \cr}\pmatrix{\Bu_r \cr \Bu_s},
\eeq{2.9a}
describing the response of a spring, where $\BF_{r}^s$ and 
$\BF_{s}^r=-\BF_{r}^s$ are the forces
node $r$ and node $s$, respectively, exert on the spring joining these nodes,
[which is the opposite of the definition given in \citeAPY{Milton:2008:RRM}],
$\Bu_r$ and $\Bu_s$ are the displacements at these two nodes, $\Bt_{rs}$ is the unit
vector pointing from node $r$ to node $s$, and $k_{rs}$ is the spring constant. Note
that the matrix entering both relations \eq{2.9} and \eq{2.9a}
is real, symmetric, degenerate and positive
semidefinite. Also $i\Go J$
is playing the role of a force, and $1/\Gm_{qrs}$ is playing the role of the elastic
spring constant (to within a proportionality factor) 
as might be expected by comparing \eq{0.1} and \eq{0.3}.
\begin{figure}
\vspace{2in}
\hspace{1.0in}
{\resizebox{2.0in}{1.0in}
{\includegraphics[0in,0in][6in,3in]{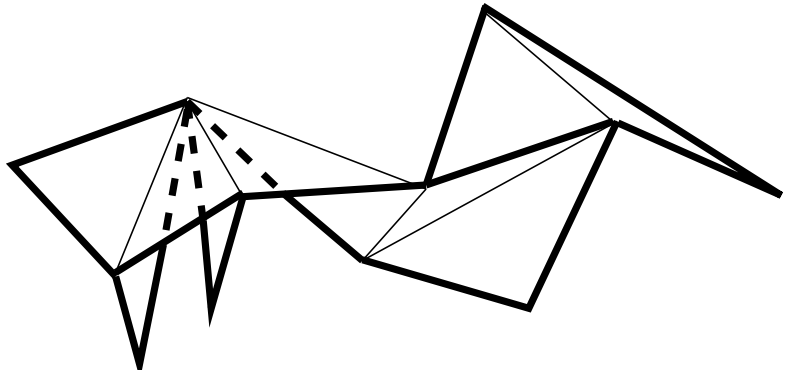}}}
\vspace{0.1in}
\caption{A magnetic circuit (M-circuit) is obtained by joining together a collection of triangular 
prisms, of the type illustrated in figure \fig{3}. At each edge there is a
small diameter cylinder, not illustrated, having $\Gve=\Gm=0$. Here 
the terminal edges are marked by thicker lines. An EM-circuit is obtained 
when a selection of the cylinders along the edges are
assigned are a dielectric constant $\Gve\ne 0$.}
\labfig{4}
\end{figure}

A magnetic circuit (the analog of a elastic network with 
springs but no masses), as illustrated in figure \fig{4}, is a 
collection of such triangular prisms, 
joined at common edges by cylinders 
having $\Gm=\Gve=0$ and with a constant diameter $d$ of the
order of $h$. In fact it is desirable to take $\Gve$ 
in these cylinders arbitrarily
small but non zero, since then $\BE$ will be constant along the
cylinder because $\BD$ is (essentially) constant. 
Edges in such a magnetic 
circuit (M-circuit)
play the role of nodes in an elastic network, and just as applied forces are 
confined to the terminal nodes in an elastic network, so too can applied free currents be 
confined to a subset of the edges in a magnetic circuit. We call these the terminal
edges, and we call the others internal edges. If a magnetic circuit contains an internal edge $qr$
which is connected to only one triangle $qrs$, then 
$T_{qr}^s=T_{rs}^q =T_{sq}^r=0$ and the triangle $qrs$ can be removed without effecting the 
response of the network. (Analogously, if a spring network contains an internal node with
only one spring and no mass attached to it then that spring can be removed without affecting the response
of the network). Thus we can restrict attention to magnetic circuits where
all internal edges are connected to at least two triangles.

Consider an internal edge  $qr$ where $m$ triangles meet at a cylinder.
Since  $T_{qr}^s/(i\Go h)$ is the value of the constant magnetic field $\BH$ 
within the triangular prism $qrs$, 
at an internal edge  $qr$ where $m$ triangles meet at a cylinder, we have
\beq \sum_{s=1}^m T_{qr}^s=0, 
\eeq{2.10}
as follows from Ampere's circuital law  
that the line integral of $\BH$ around the cylinder is zero. 
Equation \eq{2.10} is analogous to the balance of forces at a node 
in a spring network: the sum of all free currents must be zero
if there is no net free current.
At a terminal edge $qr$, Ampere's circuital law implies
\beq \sum_{s=1}^m T_{qr}^s=A_{qr},
\eeq{2.11}
where $A_{qr}/(i\Go)$ is the free current applied to that edge.

We label the edges in the network so that no edge is repeated twice, i.e. if $qr$ labels
an edge in our list, then the label $rq$ does not appear in the list. This essentially
assigns an arrow (from $q$ to $r$) to each edge, and it may be impossible to assign arrows so that
no two arrows point to the same vertex in every triangle in the circuit. Accordingly,
for example, we may want the relation \eq{2.9} to involve  $T^q_{sr}$ and $V_{sr}$ 
rather than $T^q_{rs}$ and $V_{rs}$ when the label $rs$ does not occur in the list. 
To eliminate the unwanted variables in \eq{2.9} we can then use the relations
\beq T^q_{rs}=-T^q_{sr},\quad V_{rs}=-V_{sr}, \eeq{2.11a}
which hold for all $r$, $s$ and $q$. Thus \eq{2.9} becomes
\beq\pmatrix{T_{qr}^s \cr T_{sr}^q \cr T_{sq}^r}\equiv 
\pmatrix{i\Go J_{qr}^s \cr i\Go J_{rs}^q \cr i\Go J_{sq}^r}
                          =k_{qrs}\pmatrix{1 & -1 & 1\cr
                                           -1 & 1 & -1 \cr
                                           1 & -1 & 1 \cr}
\pmatrix{V_{qr} \cr V_{sr} \cr V_{sq}},
\eeq{2.11b}
and still involves a real, symmetric, degenerate, 
positive semi-definite matrix.

To obtain an electromagnetic circuit from a magnetic circuit (including 
those magnetic circuits where some internal edges are only
connected to one triangle) we assign
a non-zero value to the dielectric constant (to some or all)
of the cylinders, of diameter $d$,  at the junctions of the triangles. 
(This is analogous to adding mass to the nodes of a spring network) 
An example is illustrated in figure \fig{4a}.
At any vertex where two or more dielectric cylinders meet we need to
make sure there is a good electrical connection between the dielectric
cylinders to allow displacement current to flow between the cylinders.

\begin{figure}
\vspace{2in}
\hspace{1.0in}
{\resizebox{2.0in}{1.0in}
{\includegraphics[0in,0in][6in,3in]{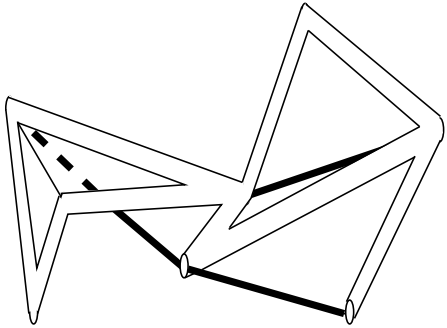}}}
\vspace{0.1in}
\caption{An electromagnetic circuit (EM-circuit) has
small diameter dielectric cylinders with $\Gve\ne 0$ and $\Gm=0$ 
along a selection of edges including possibly the terminal edges. 
Here there are three terminal edges
are marked by thicker lines.  The dielectric cylinders should be 
much thinner than drawn here. 
One internal edge, marked by the thin line, has a cylinder 
with $\Gve=\Gm=0$ attached to it.}
\labfig{4a}
\end{figure}

Now at an internal edge $qr$ where $m$ triangles meet
at a dielectric cylinder
the junction locally looks similar to the junction in
a transverse electric EM circuit where $m$ plates
meet at a dielectric cylinder, and so one expects
an equation similar to \eq{1.10} to hold.
Ampere's circuital law (with Maxwell's correction)
taken around a circuit surrounding the cylinder $qr$ implies 
\beq \sum_{s=1}^m T_{qr}^s=\Go^2 g_{qr}V_{qr}, \eeq{2.12}
where $g_{qr}=\pi d^2\Gve_{qr}/(4\ell_{qr})$, in which $\Gve_{qr}$ is
the dielectric constant of the cylinder. The term on the right arises
from the fact that $-i(\pi d^2/4)\Go\Gve_{qr}E_{qr}$ is the total displacement current
flowing through the dielectric cylinder. Inside the cylinder $\BD=\Grad\chi$
(since $\Div\BD=0$ and $\Curl\BE=0$) where $\Bn\cdot\Grad\chi=0$ at the 
cylinder walls (since $\BD=0$ in the matrix and in the triangular elements).
At the cylinder ends one has some flux of $\BD$. From the solution
to this Neumann problem $\BD$ will be essentially constant inside the
small diameter cylinder away from the ends. 
This justifies our assumption that the electric field
takes constant values along the edges of a magnetic triangular element,
at least when there are dielectric cylinders along each of these edges.

The equation \eq{2.12} which is 
essentially the same as \eq{1.10} when $I_j=0$, is the analog 
in an elastic network of Newton's law,
\beq \sum_{s=1}^m\BF_r^s=\Go^2 m_r\Bu_r, \eeq{2.13}   
describing the motion of a mass $m_r$ at a node $r$ where $m$ springs meet. At
a terminal edge \eq{2.12} generalizes to
\beq \sum_{s=1}^m T_{qr}^s=A_{qr}+\Go^2 g_{qr}V_{qr} \eeq{2.14}
where again $A_{qr}/(i\Go)$ is the free current applied to that edge.
The equations \eq{2.9}, \eq{2.12}, and \eq{2.14} hold for each 
triangle and each edge, and provide a system of equations which
can be solved to determine the response of an arbitrary electromagnetic
circuit. This will be done in section 5.

The mathematical idealization of an electromagnetic circuit
is obtained by taking the limits $h\to 0$ and $d\to 0$, while
say keeping the ratio $d/h$ fixed. The moduli of the constituent
materials need to be scaled in such a way that the parameters
entering the final equations, such as $k_{qrs}$ and $g_{qr}$,
remain fixed. Thus one should take $\Gm_{qrs}$ proportional
to $h$ (and thus very small) and $\Gve_{qr}$ proportional to
$1/d^2$ (and thus very large).

\section{Acting upon an electromagnetic circuit and creating virtual free
currents}
\setcounter{equation}{0}
One might ask how one could conceivably act on an EM-circuit, and measure
its response. A possible scenario, as sketched in figure \fig{4b},
might be to be to have electromagnetic fields incident on a body, say a cube, 
of material with $\Gm=\infty$ and $\Gve=0$ 
containing an EM-circuit, with no two terminal edges sharing a common
vertex, positioned in such
a way that only the terminal edges are exposed at the surface of the 
cube.
Let us suppose that there are no dielectric cylinders attached to
the terminal edges. Then the electric field will not be constant 
along each terminal edge. If $qr$ is a terminal edge between points
$q$ and $r$ both on the same face of the cube, then 
Faraday's law of induction implies that
the line integral
of $\BE$ along that terminal edge will play the role of the quantity
$V_{qr}$ in the electromagnetic circuit so that \eq{2.9} remains 
satisfied.

\begin{figure}
\vspace{2in}
\hspace{1.0in}
{\resizebox{2.0in}{1.0in}
{\includegraphics[0in,0in][3in,1.5in]{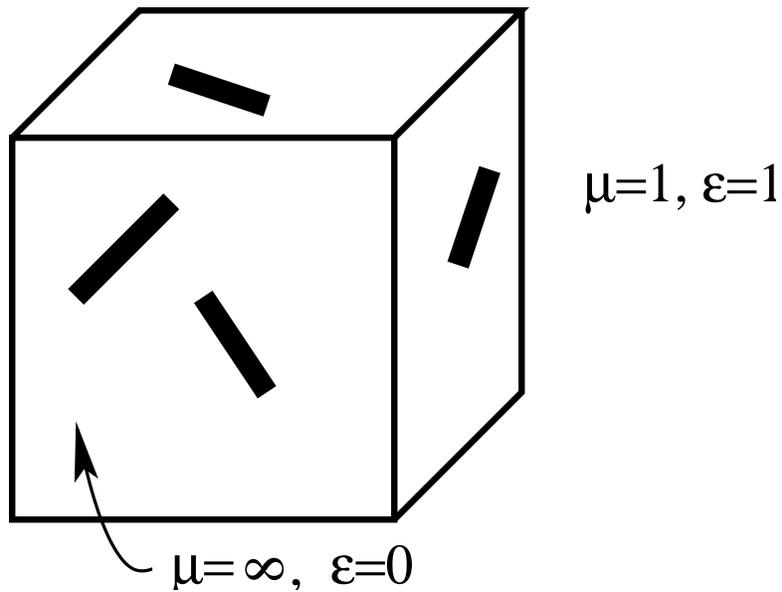}}}
\vspace{0.1in}
\caption{A cube of material with $\Gm=\infty$ and $\Gve=0$ containing
an EM-circuit with four terminal edges exposed on the 
faces of the cube. From the outside it will look as if there are
free currents flowing along the terminal edges (although in reality
they do not exist) with the endpoints acting as sources and sinks
for the displacement current field outside the cube. These virtual free currents will
be generated according to the values of the line integrals of 
$\BE$ along the terminal edges, and according to the response
matrix of the EM-circuit. In the dual setting, it 
will look like a ME-circuit generates virtual magnetic monopole currents.}
\labfig{4b}
\end{figure}

The EM-circuit causes the magnetic field $\BH$ outside 
the body to be altered in such a way that Ampere's circuital law
(with Maxwell's corrections) holds around each terminal edge.
If one was not aware of the existence of the EM-circuit,
from outside the body it would look as if the magnetic field $\BH$
near the terminal edge $qr$ was generated by
a free-current $-A_{qr}/(i\Go)$
flowing from $q$ to $r$. In other words, if one incorrectly assumes
that $\BH=0$ throughout the cube, then Ampere's circuital law
would falsely imply the existence of this free current,
which we call a virtual free current, flowing along the terminal edge.
It is nothing else but the surface free current which the EM-circuit
exerts on the surrounding material at  the terminal edge $qr$.
(In a similar fashion one can insert a mass-spring network into
a cavity in an elastic body, with only the terminal nodes attached
to the boundary of the cavity. If one is not aware of the existence of 
the spring mass network from outside the cavity, it would look
like the stress field in the body was altered by concentrated
forces acting at the positions of the terminal nodes.)    

Now the internal edges will carry some displacement current $A_{qr}/(i\Go)$
out of the vicinity of the point $q$ and a displacement current 
$A_{qr}/(i\Go)$ into  the vicinity of $r$. If one is not aware of 
the existence of the electromagnetic circuit it would look like
the point $p$ is a current source and the point $r$ is a current sink:
it would look like the ends of the virtual free-current $-A_{qr}/(i\Go)$
along the terminal edge, act as sources and sinks for the displacement
current outside the body. 

If the thickness $h$ of each terminal edge is very small, then
the coupling between the electromagnetic circuit and the 
fields in the exterior will be weak. As in the case
of transverse electric EM-circuits, small virtual free-currents
along the terminal edges will cause the field $\BE(\Bx)$ to
be modified in the near vicinity of each terminal edge. 
One suggestion to enhance the 
coupling is cap each terminal edge $qr$ with a 
$\Gve=\Gm=0$ semicircular cylinder of length $\ell_{qr}$ and 
diameter $d_0$, where $d_0$ is not small. At the two ends 
of this cylinder one could attach $\Gve=\infty$, $\Gm=0$
quarter spheres of diameter $d_0$, to allow the displacement current to 
enter and exit the points $q$ and $r$ with little resistance.
In these quarter spheres $\BE=0$. Faraday's law of induction
then implies that the line integral of $\BE$ along the outer surface of
the semicircular cylinder will equal the line integral of $\BE$
along the terminal edge.

\section{A formula for the response matrix of an EM-circuit, and the
properties of this response matrix}
\setcounter{equation}{0}
In a magnetic or electromagnetic circuit with $n$ terminal edges let us suppose
these edges have been numbered from $1$ to $n$. Then the response of the network
is governed by the linear relation
\beq \BA=\BW\BV \eeq{2.15}
between the terminal variables $\BA=(A_1,A_2,\ldots,A_n)$ which measure the 
real or virtual free
currents at these edges, and the variables $\BV=(V_1,V_2,\ldots,V_n)$ which measure
the line integral of the electric field along these edges. 
When all the edges in the circuit
are terminal edges the response matrix $\BW$ equals a symmetric matrix $\BW^0$ with an
especially simple form. From \eq{2.9} and \eq{2.14} the diagonal elements of
$\BW^0$ are given by
\beq W^0_{qr,qr}=-\Go^2 g_{qr}+\sum_{s}k_{qrs}, \eeq{2.16}
where the sum is over vertices $s$ such that $qrs$ is a triangle in the circuit,
while the off-diagonal elements $W^0_{qr,st}$ are zero when $qr$ and $st$ are not
two edges of some triangle in the circuit, and the remaining off diagonal
elements are each given by one of the formulas
\beq W^0_{qr,rs}=W^0_{qr,sq}=k_{qrs}, \quad  W^0_{qr,sr}=W^0_{qr,qs}=-k_{qrs},
\eeq{2.17}
according to what edge labels are in our list, where $qrs$ is
a triangle in our circuit. Suppose we divide these
edges into two groups, and order the edges so that one group comes first.
Then the matrix relation \eq{2.15} takes the block form
\beq \pmatrix{\BA_1 \cr \BA_2}=\pmatrix{\BW^0_{11} & \BW^0_{12}\cr
                                       (\BW^0_{12})^T & \BW^0_{22}}
\pmatrix{\BV_1 \cr \BV_2},
\eeq{2.18}
where $\BA_1$ and $\BV_1$ are the set of variables associated with the
first group and $\BA_2$ and $\BV_2$ are the set of variables associated with the
second group. Now consider the case where the first group are terminal edges,
while the second group are internal edges. Then $\BA_2=0$ and \eq{2.18}
implies $\BA_1=\BW\BV_1$ with the response matrix $\BW$ 
of the circuit being the Schur complement
\beq \BW=\BW^0_{11}-\BW^0_{12}(\BW^0_{22})^{-1}(\BW^0_{12})^T.
\eeq{2.19}
This is our formula for the response matrix of an arbitrary 
electromagnetic circuit. In particular it shows that the 
response matrix is always symmetric. It
may be that the matrix $\BW^0_{22}$ is singular, in which case
if $\BA_1$ is finite there are generally restrictions on the possible values
that $\BV_1$ can take. 

Also recall that the matrix entering the relation \eq{2.9} is positive
semidefinite. Therefore if $\Gm_{qrs}$ has a non-negative 
imaginary part, and hence $k_{qrs}$ has a non-positive imaginary 
part, for each triangle $qrs$ in the circuit and $\Gve_{qr}$, and 
hence $g_{qr}$, have a non-negative imaginary part for each edge $qr$ in the 
circuit, the imaginary part of $\BW^0$ will be negative semidefinite, being
a sum of negative semidefinite matrices. It follows that the quantity
\beq S=\BA'\cdot\BV''-\BA''\cdot\BV'
=-\BV'\cdot(\BW^0)''\BV'
-\BV''\cdot(\BW^0)''\BV''
\eeq{2.20}
will be non-negative, where the primes denote real parts, and 
the double primes imaginary parts.
In particular if $\BA_2$ in \eq{2.18} is zero,
the left hand side of the above equation reduces to
\beq  S=\BA_1'\cdot\BV_1''-\BA_1''\cdot\BV_1'
=-\BV_1'\cdot\BW''\BV_1'
-\BV_1''\cdot\BW''\BV_1'',
\eeq{2.21}
and since this is non-negative for all values of $\BV_1$ we deduce that
$\BW''$, like $(\BW^0)''$, is negative semidefinite.
\section{The energy stored and dissipated in an electromagnetic circuit}
\setcounter{equation}{0}
To obtain a formula for the time averaged 
energy stored in an electromagnetic circuit
let us assume the moduli in the electromagnetic circuit are real and
do not depend on frequency. Recall that the physical electric and
magnetic fields are the real part of $\BE e^{-i\Go t}$ and 
$\BH e^{-i\Go t}$. Locally the time averaged electric and magnetic field 
energy densities will therefore be $\Gve|\BE|^2/4$ and $\Gm|\BH|^2/4$.
In the magnetic element $qrs$ discussed at the beginning of section 3
the time averaged stored magnetic energy will be
\beq ha_{qrs}\Gm_{qrs}|H_3|^2/4=|T_{qr}^s|^2/(4\Go^2k_{qrs}),
\eeq{2.22}
while in the dielectric cylinder $qr$ the time averaged 
stored electrical energy will be
\beq (\pi\ell_{qr}d^2/4)\Gve_{qr}|E_{qr}|^2/4=|V_{qr}|^2 g_{qr}/4.
\eeq{2.23}
The total electromagnetic energy stored in the electromagnetic
circuit will be a sum of such expressions taken over all
magnetic elements and dielectric cylinders in the circuit.
An appealing feature is that the resultant expression only
depends on $\Go$ and the
parameters $T$, $V$, $k$ and $g$ characterizing
the electromagnetic circuit, and not on the parameters $a_{qrs}$,
$h$, $\ell_{qr}$, and $d$.

Now let us consider how much electromagnetic energy is dissipated
into heat within the electromagnetic circuit when the moduli 
are complex and depend on frequency. Locally the time averaged electrical
and magnetic power dissipated into heat per unit volume
will be $\Go\Gve''|\BE|^2/2$ and $\Go\Gm''|\BH|^2/2$, respectively. Within
the magnetic element $qrs$ this will integrate to
\beqa \Go ha_{qrs}\Gm_{qrs}''|H_3|^2/2 & = & (1/k_{qrs})''|T_{qr}^s|^2/(2\Go)
\nonum
& = & [(T_{qr}^s)'(T_{qr}^s/k_{qrs})''-(T_{qr}^s)''(T_{qr}^s/k_{qrs})']/(2\Go).
\eeqa{2.24}
Now we can substitute \eq{2.9} into this, and associate a portion 
of the resultant
expression to each edge, where the portion assigned to edge $qr$ is
\beq [(T_{qr}^s)'V_{qr}''-(T_{qr}^s)''V_{qr}']/(2\Go). \eeq{2.25}
In the dielectric cylinder $qr$ the time averaged electrical
power dissipated into heat is
\beqa (\pi\ell_{qr}d^2/4)\Go\Gve_{qr}''|E_{qr}|^2/2& = & 
\Go|V_{qr}|^2 g_{qr}''/2 \nonum & = & 
[(-\Go^2 g_{qr}V_{qr})'V_{qr}''-(-\Go^2 g_{qr}V_{qr})''V_{qr}']/(2\Go).
\eeqa{2.26}
Adding up all the contributions \eq{2.25} and \eq{2.26}
associated with edge $qr$ and using the relation \eq{2.14} we see
that the total contribution associated with edge $qr$ is 
zero for an internal edge and
\beq [A_{qr}'V_{qr}''-A_{qr}''V_{qr}']/(2\Go)
\eeq{2.27}
for a terminal edge. By summing this expression over all terminal
edges we see that the quantity $S/(2\Go)$, where $S$ is given by \eq{2.21},
is the time averaged electromagnetic energy converted into heat
in the circuit. 

This is consistent with Poynting's theorem. Suppose we attached to
the edge $qr$ a rectangular plate of thickness $h$ and width $\ell_{qr}$
in which there is a magnetic field $\BH$ with component 
$H_{qr}=-A_{qr}/(i\Go h)$ perpendicular to the plate (and

surrounded by material with $\Gm=\infty$ and $\Gve=0$) so that
Ampere's circuital law (with Maxwell's corrections) is 
satisfied around the terminal edge. At any instant in time
the flux of energy into the terminal $qr$ will be 
$h\ell_{qr}(E_{qr}e^{-i\Go t})'(H_{qr}e^{-i\Go t})'$,
so the time averaged energy flux is
\beq h\ell_{qr}(E_{qr}'H_{qr}'+E_{qr}''H_{qr}'')/2
=(-V_{qr}'A_{qr}''+V_{qr}''A_{qr}')/(2\Go).
\eeq{2.28}
Thus the quantity \eq{2.27} has the physical interpretation
as this time averaged energy flux, and it is then natural
that its sum over all terminal edges should be  
the time averaged electromagnetic energy converted into heat
in the circuit.

\section{A correspondence between electrical circuits and a subclass of electromagnetic circuits}
\setcounter{equation}{0}
At fixed frequency, linear electrical circuits correspond to a subclass of EM-circuits,
namely those where there are a sufficient number of magnetic elements 
and these all have $\Gm_{qrs}=0$.
Let us consider, for simplicity, an $n$-terminal planar electrical
network with terminal nodes at the vertices of a polygon and with the remainder
of the circuit lying with the polygon. If $J_{qr}$ is the complex current flowing
from node $q$ to node $r$ and these nodes have complex voltages $V_q$ and $V_r$, then we 
have
\beq J_{qr}=Y_{qr}(V_q-V_r), \eeq{3.0}
where  $Y_{qr}$ is the complex admittance (having non-negative real part) of the
circuit element joining these two nodes.

\begin{figure}
\vspace{2in}
\hspace{0.0in}
{\resizebox{2.0in}{1.0in}
{\includegraphics[0in,0in][6in,3in]{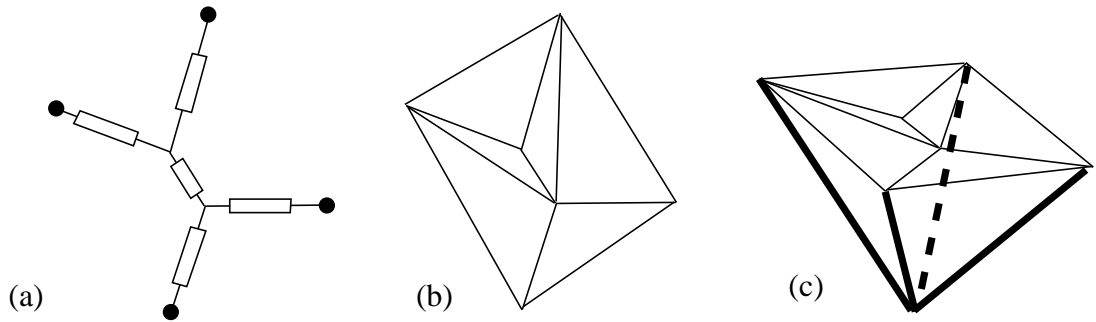}}}
\vspace{0.1in}
\caption{Construction of an EM-circuit corresponding to the planar electrical 4-terminal
network (a). The first step in (b) is to triangulate the network, and place appropriately
valued dielectric cylinders (not shown) along the edges, and magnetic triangles with $k=\infty$
in each triangle. Then one adds a vertex below the network, and magnetic
triangles with $k=\infty$ on the four triangular sides. The four new edges, marked 
by thicker lines, are the terminal edges of the EM-circuit.}
\labfig{5}
\end{figure}

For example, one may consider the
four terminal network of figure \fig{5}(a) which has two internal nodes.
To build an associated EM-circuit, the first step is to triangulate the network by adding additional edges
with zero admittance, as illustrated in \fig{5}(b). To each triangle
formed by this triangulation (not containing any nodes)
with vertices $q$, $r$ and $s$ we assign a constant $k_{qrs}=\infty$.
In the limit as $k_{qrs}\to \infty$ the equation \eq{2.9} reduces to 
\beq T_{qr}^s  = T_{rs}^q=T_{sq}^r, \quad V_{qr}+V_{rs}+V_{sq}=0.
\eeq{3.1}
Following the ideas of \citeAPY{Engheta:2005:CEO} and \citeAPY{Engheta:2007:CLN}
we attach to each edge $qr$ a dielectric cylinder with constant $g_{qr}=iY_{qr}/\Go$,
which therefore will have non-negative imaginary part. [If the circuit element is
a capacitor, then this will correspond to taking a value of the dielectric 
constant $g_{qr}$ which is real and positive; if the circuit element is 
a resistor, then this will correspond to taking $g_{qr}$ with zero real part
and positive imaginary part; if the circuit element is 
an inductor, then this will correspond to taking $g_{qr}$ almost real
and negative.]

The equation \eq{2.12} then becomes
\beq J_{qr}=Y_{qr}V_{qr}, \quad {\rm where}~J_{qr}=\frac{1}{i\Go}\sum_{s=1}^mT^s_{qr}=\sum_{s=1}^mJ^s_{qr},
\eeq{3.2}
in which $m=1$ or $2$  is the number of triangles sharing the edge $qr$, and $s$ indexes
each of these triangles.

We next introduce an additional node 0 below the network, and for
each pair $t$ and $u$ of neighboring terminal nodes around the polygon we construct 
the triangle $tu0$ with constant $k_{tu0}=\infty$. As illustrated in figure \fig{5}(c).
This implies we have
\beq  T_{tu}^0= T_{u0}^t=T_{0t}^u, \quad V_{tu}+V_{u0}+V_{0t}=0.
\eeq{3.3}
The edges $u0$, with $u=1,2,\ldots n$  are taken as the terminal edges of the electrodynamic circuit,
and no dielectric cylinders are attached to them. The second equations in 
\eq{3.1} and \eq{3.3} imply that we can assign a voltage $V_q$ to each 
node such that
\beq V_{qr}=V_q-V_r, \quad V_0=0,\quad V_{u0}=V_u.
\eeq{3.4}
Thus \eq{3.2} reduces to \eq{3.0}. Also the first equations in \eq{3.1} and \eq{3.3}
ensure that the total current is a sum of loop currents. Therefore Kirchoff's 
law that the sum of currents flowing into a node equals the sum of currents flowing
out of that node is automatically satisfied. Thus the standard electrical circuit
equations are satisfied. 

Now the terminal edge variables $V_{u0}$, $u=1,2,\ldots n$,  are the voltages at the
terminal nodes of the electrical circuit. Also it is easy to see that
the terminal edge variable $J_{u0}=A_{u0}/(i\Go)$ is the net current
flowing out of the electrical circuit from node $u$ to node $0$.
Thus the map $\BW/(i\Go)$ is the Dirichlet to Neumann map of the electrical
circuit. 

If the electrical circuit is non-planar, then we modify the circuit so that
all the terminal nodes are at the vertices of a (not necessarily convex) polygon
lying on a plane below the circuit. Then the circuit
above the plane is appropriately triangulated by adding additional nodes 
if necessary. A magnetic element with $k=\infty$ is inserted in each triangle
and appropriately valued dielectric cylinders are attached to the edges. Each pair
of neighboring terminal nodes on the polygon are then attached with a magnetic element 
having $k=\infty$ to an additional ground node situated below the plane. Each 
edge between a terminal node and the ground node is a terminal edge of the
resulting EM-circuit.

\section{Electromagnetic ladder networks, a characterization of their possible
response matrices, 
and a material with non-Maxwellian
macroscopic behavior}
\setcounter{equation}{0}
Electromagnetic circuits can have many different topologies and seems
very difficult to characterize their possible macroscopic matrices 
$\BW$, i.e. classify (for a given topology of terminal edge connections?)
which matrices are realizable as the Schur complement of a matrix $\BW^0$
with elements \eq{2.16} and \eq{2.17}, and which ones not.
Here we restrict our attention to an important subclass of electromagnetic circuits, called electromagnetic ladder networks (EM ladder networks),
for which such a characterization is possible.  Consider, as illustrated in 
the simple EM-circuit consisting of
two magnetic triangles $qrs$ and $rst$ joined by  a cylinder with $\Gve=\Gm=0$ 
along the internal edge $rs$.
Assume they have the same constant $k_{qrs}=k_{rst}=k$. Then \eq{2.9}
implies
\beqa T_{qr}^s & = & T_{rs}^q=T_{sq}^r=k(V_{qr}+V_{rs}+V_{sq}), \nonum
 T_{rs}^t & = & T_{st}^r=T_{tr}^s=k(V_{rs}+V_{st}+V_{tr}).
\eeqa{4.1}
The edges $sq$ and $tr$, labeled 1 and 2, are taken to be the terminal edges. They are without
dielectric cylinders, so \eq{2.14} and \eq{2.12} imply
\beq T_{sq}^r=A_{sq}\equiv -I_{12},\quad T_{tr}^s=A_{tr}\equiv -I_{21}, \quad  T_{rs}^q+T_{rs}^t=0.
\eeq{4.2}
Suppose there are dielectric cylinders along the internal edges $qr$
and $st$ with the same constants $g_{qr}=g_{st}=g$. At these 
edges \eq{2.12} implies
\beq T_{qr}^s=\Go^2g V_{qr}, \quad  T_{st}^r=\Go^2g V_{st}.
\eeq{4.3}
Solving these equations for $I_{12}$ and $I_{21}$ in terms of
$V_1\equiv V_{sq}$ and $V_2 \equiv V_{tr}$ gives 
\beq I_{12}=-I_{21}=k_{12}(V_2-V_1),
\eeq{4.4}
where
\beq k_{12}=\frac{1}{2}[1/k-1/(\Go^2g)]^{-1}
\eeq{4.5}
has non-positive imaginary part, because $k$ has non-positive imaginary 
part and $g$ has non-negative imaginary part. From now on we ignore the
internal edges of this simple EM-circuit, treating the simple EM-circuit
itself as a basic ladder network element. 

\begin{figure}
\vspace{2in}
\hspace{1.0in}
{\resizebox{2.0in}{1.0in}
{\includegraphics[0in,0in][6in,3in]{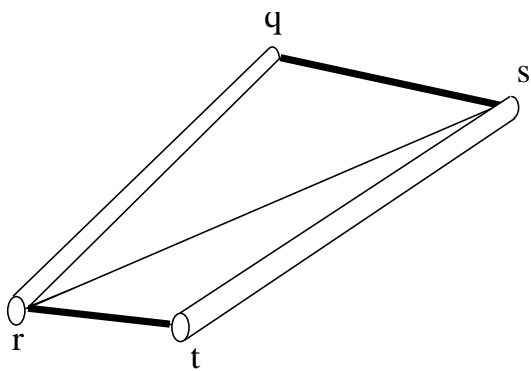}}}
\vspace{0.1in}
\caption{A simple EM-circuit which is the basic element for constructing
EM ladder networks. The triangles $qrs$ and $rst$ are magnetic elements, and
dielectric cylinders are attached to the edges $qr$ and $st$. A cylinder,
not shown, with $\Gve=\Gm=0$ is attached to the edge $rs$ to join the
two magnetic elements. The edges 
$rt$ and $qs$ (which need not be coplanar) are terminal edges.}
\labfig{6}
\end{figure}

The relation \eq{4.4} is similar to that associated with an element in an 
electrical circuit, although the physical interpretation of the
variables is completely different.
In the setting of an electrical circuit, using the notation 
of \citeAPY{Milton:2008:RRM}, 
1 and 2 label two nodes, $V_1$ and $V_2$ are the 
potentials at these nodes, $iI_{12}/\Go$ is the current flowing
from node 1 to node 2, while $iI_{21}/\Go$ is the current flowing
from node 2 to node 1, and $k_{12}=1/L$ for an inductor,
$k_{12}=-\Go^2C$ for a capacitor, and $k_{12}=-i\Go/R$ for a resistor,
where $L$ is the inductance, $C$ the capacitance, and $R$ the resistance.

Building upon this analogy we can join a set of these simple EM
circuits together, to obtain what we call an EM ladder network,
as illustrated in figure \fig{7}(a).  An $n$-terminal
EM ladder network consists of $n+m$ edges $Q_\Ga$ indexed by $\Ga=1,2,\ldots, n+m$,
having no vertex in common. Each pair of edges $(Q_\Ga, Q_\Gb)$ may have a simple 
EM-circuit (of the type just discussed) joining them, and from \eq{4.4}, we 
have the relation
\beq  I_{\Ga\Gb}=-I_{\Gb\Ga}=k_{\Ga\Gb}(V_\Gb-V_\Ga),
\eeq{4.6}
where $k_{\Ga\Gb}$ is the constant associated with the simple EM-circuit,
and $k_{\Ga\Gb}=0$ if there is no simple EM-circuit joining the edges.  
The first $n$ edges  $Q_\Ga$
are the terminal edges
of the EM ladder network (not to be confused with the terminal edges of the simple
EM-circuits, which are all the edges $Q_\Ga$), and the remaining $m$ edges are
internal edges.

\begin{figure}
\vspace{2in}
\hspace{1.0in}
{\resizebox{2.0in}{1.0in}
{\includegraphics[0in,0in][6in,3in]{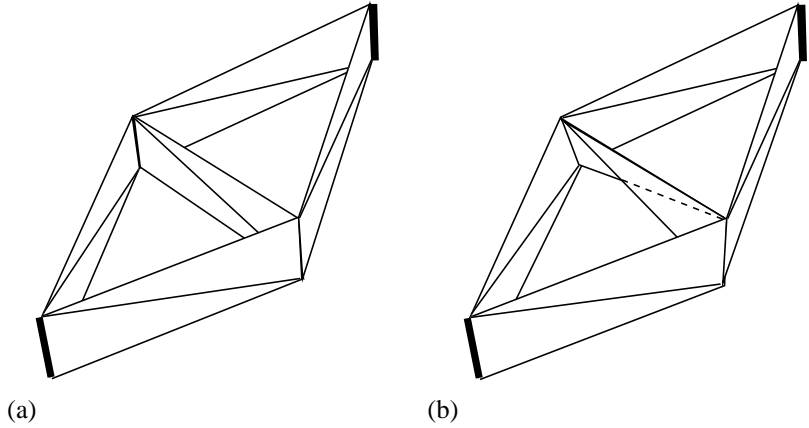}}}
\vspace{0.1in}
\caption{Figure (a) shows an EM ladder network which is the analog
of the classical Wheatstone bridge, although its physical
behavior is completely different. 
The two terminal edges are marked by the thicker lines,
and for simplicity the dielectric cylinders along the edges of each basic
ladder network element are not shown. Dielectric cylinders could be inserted 
along the edges between neighboring basic ladder network elements, in which case
the EM ladder network becomes equivalent of a Wheatstone bridge with each node
connected by a capacitor to ground. In figure (b) we have twisted the 
central bridge element so that it has a response governed by \eq{4.9}}
\labfig{7}
\end{figure}

Each pair of edges $(Q_\Ga, Q_\Gb)$ may have an elementary 
EM-circuit of the type just discussed joining them. Each edge $Q_\Ga$,
may have a dielectric cylinder, with constant $g_\Ga$ attached to it.
If this edge is an internal edge of the EM ladder network then from \eq{2.12} we have
\beq \sum_{\Gb=1}^{n+m}I_{\Ga\Gb}=-\Go^2g_\Ga V_\Ga, \eeq{4.7}
(in which we set $I_{\Ga\Ga}=0$)
while if this edge is a terminal edge of the EM ladder network then from \eq{2.14} we have, 
\beq A_\Ga+\sum_{\Gb=1}^{n+m}I_{\Ga\Gb}=-\Go^2g_\Ga V_\Ga. 
\eeq{4.8}
The response of the EM ladder network is then governed by the relation
$\BA=\BW\BV$ between the terminal variables $\BA=(A_1,A_2,\ldots,A_n)$ which measure the free
currents at these edges, and the variables $\BV=(V_1,V_2,\ldots,V_n)$ which measure
the tangential electric field at these edges.

Equations \eq{4.6}-\eq{4.8} are the same as those for electrical circuit
in which the nodes may be connected to ground by a capacitor. It then
follows directly from the results of \citeAPY{Milton:2008:RRM} that for any
fixed real frequency $\Go$ any real symmetric matrix $\BW$ may be realized
by an EM ladder network having real positive values of the constants $k_{\Ga\Gb}$
and $g_\Ga$, and any complex symmetric matrix $\BW$ with ${\rm Im}\BW\geq 0$
can be realized by an EM ladder network having real positive values of the constants $k_{\Ga\Gb}$
and complex values of the constants $g_\Ga$ having non-negative real and imaginary
parts. Thus at fixed frequency we have a complete characterization of the possible response matrices $\BW$ 
of EM ladder networks, both in the lossless case, and in the lossy case. We also have a complete
characterization of the possible response matrices $\BW$ associated with the class of electromagnetic circuits
where no two terminal edges are connected, since an EM ladder network can be constructed having these
edges as its terminals, and having the desired response matrix $\BW$.

We can introduce another basic ladder network element of an EM ladder network. Just by reversing the roles of the vertices $t$ and $r$
in the original basic ladder network element and using \eq{2.11a}, we obtain a basic ladder network element with the response
\beq I_{12}=I_{21}=-k_{12}(V_1+V_2).
\eeq{4.9}
Of course utilizing such basic ladder network elements, as done in the example of figure \fig{7}(b),
does not enlarge the class of possible response matrices $\BW$ 
of EM ladder networks at fixed frequency (which is already as large as possible without the introduction of such
elements). 

Let us now sketch how one could get a material with non-Maxwellian
macroscopic behavior using electromagnetic ladder networks.
In the same way that one can build a cubic network of resistors so too
can one build a cubic EM ladder network of basic network elements with the 
response \eq{4.6}
joined at edges $Q_\Ga$, with no dielectric cylinders attached to 
these edges (so that all $g_{\Ga}=0$ and it corresponds to the cubic network of resistors). Just as 
the cubic network of resistors responds macroscopically as a material
with some effective conductivity, so too will the cubic EM ladder network 
respond macroscopically in a way which is definitely non-Maxwellian. 
Without being too specific, for
a periodic ladder network one will have some relation of the form $\BK=\BGS\Grad V$,
where $V(\Bx)$ is a suitably scaled local average of the variables $V_\Ga$,
$\BGS$ is the matrix governing the effective response,
and $K_i(\Bx)$, $i=1,2,3$, is a suitably scaled local average of the variables $I_{\Ga\Gb}$
taken over the subset of basic ladder network elements which are ``aligned'' parallel
to the $x_i$-axis. If such a cubic EM ladder network is embedded in a large cube 
having $\Gm=\infty$ and $\Gve=0$ with the terminal edges exposed at the
boundary of the cube, then the interface conditions between the electromagnetic
fields outside the cube, and the fields $\BK(\Bx)$ and $V(\Bx)$ inside the cube will
presumably depend on the geometric microconfiguration of the terminal
edges of the EM ladder network at the cube faces. Obviously there is much to explore
here.

\section*{Acknowledgements}
We are grateful to the referees for their comments. 
Graeme Milton is thankful for support from 
the Universit\'e de Toulon et du Var and from the
National Science Foundation through grant DMS-070978.
Pierre Seppecher is grateful for travel support
from the University of Utah.

\bibliography{/u/ma/milton/tcbook,/u/ma/milton/newref}

\end{document}